\documentclass[12pt,a4paper]{conference}

\usepackage{fancyhdr}
\usepackage{graphicx,amsmath,amssymb,cite}
\usepackage{multind}
\makeindex{author} \makeindex{subject}

\newcommand{\beq}{\begin{equation}}
\newcommand{\eeq}[1]{\label{#1}\end{equation}}
\newcommand{\eeqn}{\end{equation}}

\newcommand{\beqa}{\begin{eqnarray}}
\newcommand{\eeqa}[1]{\label{#1}\end{eqnarray}}
\newcommand{\eeqan}{\end{eqnarray}}

\let\bar=\overbar

\newcommand{\Dslash}{\not{\hbox{\kern-4pt $D$}}}
\newcommand{\dslash}{\not{\hbox{\kern-2pt $\del$}}}

\newcommand{\msb}{{\bar{\ssstyle M \kern -1pt S}}}

\begin{document}

\Chapter{QCD SYMMETRIES IN EXCITED HADRONS}
           {QCD symmetries in excited hadrons}{L. Ya. Glozman}
\vspace{-6 cm}\includegraphics[width=6 cm]{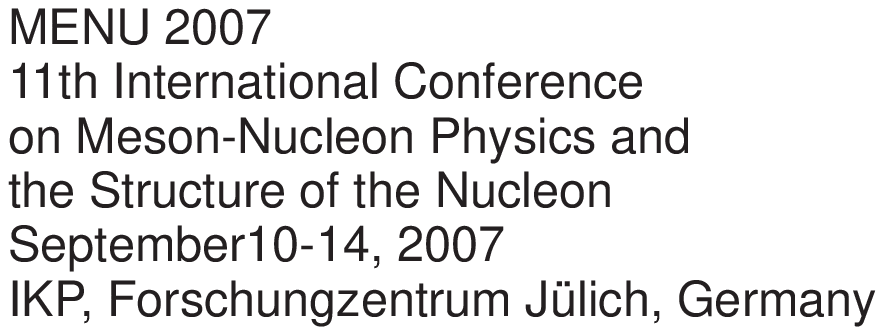}
\vspace{4 cm}

\addcontentsline{toc}{chapter}{{\it N. Author}} \label{authorStart}

\begin{raggedright}

{\it L. Ya. Glozman }\index{author}{Author, N.}\\
Institute for Physics, Theoretical Physics Branch\\
University of Graz\\
Universit\"atsplatz 5, A-8010 Graz, Austria
\bigskip\bigskip

\end{raggedright}

\begin{center}
\textbf{Abstract}
\end{center}
Recent developments for chiral and $U(1)_A$ restorations in excited
hadrons are reviewed. We emphasize predictions of the chiral symmetry
restoration scenario for axial charges and couplings to Goldstone
bosons. Using very general chiral symmetry arguments it is shown that
strict chiral restoration in a given excited nucleon forbids its decay
into the $N \pi$ channel. We confront this prediction with the 
$N^*N\pi$ coupling constants extracted from the decay widths and observe
a 100 \% correlation of these data with the spectroscopic parity doublet
patterns. These results suggest that the lowest approximate chiral parity
doublet is the $N(1440) - N(1535)$ pair. In the meson sector we discuss
predictions of the chiral symmetry restoration for still missing states
and a signature of the higher symmetry observed in new $\bar p p$
data. We conclude with the exactly solvable chirally symmetric and
confining model that can be considered as a generalization of the
1+1 dimensional 't Hooft model to 4 dimensions. Complete spectra
of $\bar q q$ mesons demonstrate a fast chiral restoration with increasing
$J$ and a slow one with increasing $n$.

\section{Introduction}

In hadrons consisting of $u$ and $d$ quarks there are two crucially
important properties of QCD - chiral symmetery and confinement. Their
interrellations and mechanisms are not yet understood. What we do know
theoretically is that at zero temperature and density in the confining
phase chiral symmetry must be necessarily spontaneously broken in
the vacuum \cite{hooft}. Another conceptual and closely related issue
is the generation of hadron mass in the light quark sector. It was
considered almost self-evident that such a mass generation proceeds
via the chiral symmetry breaking in the vacuum and the most important
characteristics that determines the hadron mass is the quark condensate
of the vacuum. Indeed, it is firmly established both phenomenologically
and on the lattice that to leading order the pion mass squared is
proportional to the bare quark mass and the quark condensate \cite{GOR}.
In the baryon sector the very absence of the chiral partner to the nucleon
implies that its mass is at least mostly  related to the spontaneous breaking
of chiral symmetry in the vacuum. This fact is supported by the Ioffe
formula \cite{ioffe} that connects, though not rigorously, the nucleon
mass with the quark condensate. Another obvious sign of the strong
dynamical chiral symmetry breaking effects in the nucleon is the
large pion-nucleon coupling constant. Indeed, it is well understood that
the coupling of the Goldstone bosons to the nucleon is a direct consequence
of the spontaneous chiral symmetry breaking and is a basis for nucleon chiral
perturbation theory \cite{bcpt}. One more strong evidence for the chiral
symmetry breaking in the nucleon is its large axial charge, $g_A = 1.26$. 

A main message
of this talk is that the mass generation mechanism in excited hadrons
is essentially different - the quark condensate of the vacuum becomes
less and less important with the excitation and the chiral as well as
the $U(1)_A$ symmetries get
eventually approximately restored in the given hadron, even though they are
strongly broken in the vacuum. This is referred to as effective restoration
of chiral symmetry, for a review see ref. \cite{G3}. 

It is important to precisely characterize what is implied under effective
restoration of chiral and $U(1)_A$ symmetry in excited hadrons. A 
mode of symmetry is defined only by the properties of the vacuum.
If a symmetry is spontaneously broken in the vacuum, then it is the
Nambu-Goldstone mode and the whole spectrum of excitations on the
top of the vacuum is in the Nambu-Goldstone mode. However, it may happen
that the role of the chiral symmetry breaking condensates becomes
progressively less important higher in the spectrum, because the
valence quarks decouple from the quark condensates. This means
that the chiral symmetry breaking effects become less and less
important in the highly excited states and asymptotically the
states approach the regime where their properties are determined
by the underlying unbroken chiral symmetry (i.e. by the symmetry
in the Wigner-Weyl mode). This effective restoration 
in excited hadrons should not
be confused with the chiral symmetry restoration in the vacuum at
high temperature/density. In the latter case the quark vacuum becomes
trivial and the system is in the Wigner-Weyl mode. In the former case
the symmetry is always broken in the vacuum, however this symmetry breaking
in the vacuum gets irrelevant in the highly excited states.

\section{Empirical
evidence for chiral restoration in excited nucleons}

The nucleon excitation spectrum is shown in Fig. 1. Only well-established
states (i.e. without stars in boxes) should be seriously considered. It is
well seen that there is no chiral partner to the nucleon. This necessarily
implies that chiral symmetry is strongly broken in the nucleon and consequently
is realized nonlinearly \cite{wein}. Obvious approximate parity doublets
are observed in the region 1.7 GeV and higher. 
 An absence of parity doublets for the
lowest-lying states and their apparent appearance for (highly) excited states
was taken in refs. \cite{G1,CG1,CG2,G2} as evidence for chiral restoration 
in excited baryons, for a review see ref. \cite{G3}.
The parity doublets in the 1.7 GeV region have
been assigned to the $(0,1/2)+(1/2,0)$ representation of the parity-chiral
group because there are no approximately degenerate doublets in the same mass
region in the spectrum of the delta-resonance \cite{CG2,G3}.
 A clear testable prediction of
the chiral symmetry restoration scenario is  an existence of  chiral partners
 of the well established high-lying resonances $N(2190)$ and $N(2600)$.
  A dedicated
 experimental search of these missing states can be undertaken \cite{S}.
Similar situation takes place in the Delta-spectrum. 

\begin{figure}[hb]
\begin{center}
\includegraphics[height=8cm,angle=-90,clip=]{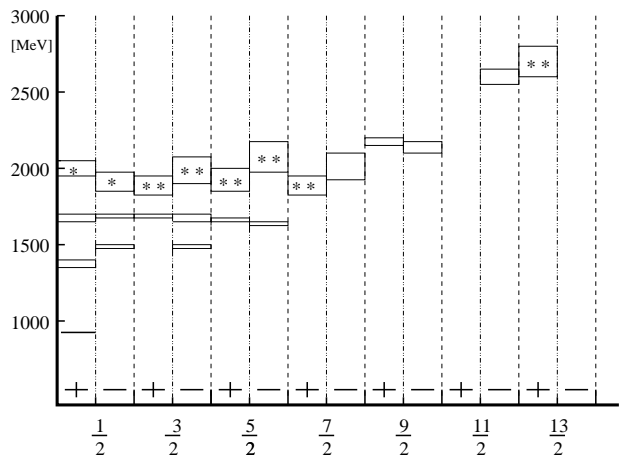}
\label{ns}
\caption{Low- and high-lying nucleons. Those states which are
not yet established are marked by ** or * according to the PDG classification.}
\end{center}
\end{figure}

While these parity doubling patterns are impressive, they are still only 
suggestive,
because so far no other complementary experimental data would independently
tell us that these parity doublets are due to effective chiral symmetry
restoration. Strict chiral restoration in a given baryon would imply that its
diagonal axial charge is zero and hence the diagonal coupling constant to 
the pion must vanish \cite{G3,CG3,GN1,JPS1,JPS2}. This is one of the most
important implications of the chiral symmetry restoration and is reviewed below.

Assume that we have a free $I=1/2$ chiral doublet $B$ in the $(0,1/2)+(1/2,0)$
representation  and there are no chiral symmetry breaking terms.  This
doublet is a column \cite{LEE}

\begin{equation}
B = \left(\begin{array}{c}
B_+\\
B_-
\end{array} \right),
\label{doub}
\end{equation}

\noindent
where the bispinors
$B_+$ and $B_-$ have positive and negative parity, respectively.
The chiral transformation law
 under the $(0,1/2) \oplus (1/2,0)$ representation 
 provides a mixing of two fields  $B_+$ and $B_-$ \footnote{Note that
 the axial transformation given in \cite{LEE} is incorrect as it
 breaks chiral symmetry of the kinetic term. The correct axial
 transformation is given in ref. \cite{G3}.}

\begin{equation}
B \rightarrow 
\exp \left( \imath \frac{\theta^a_V \tau^a}{2}\right)B; ~~
B \rightarrow 
\exp \left(  \imath \frac{\theta^a_A\tau^a}{2} \sigma_1
\right)B.
\label{VAD}
\end{equation}

\noindent
Here $\sigma_i$ is a Pauli matrix that acts in the  $2 \times 2$
space of the parity doublet. 
Then  the chiral-invariant Lagrangian of the free parity doublet is given as

\begin{equation}
\mathcal{L}_0  =  i \bar{B} \gamma^\mu \partial_\mu B - m_0 \bar{B}
B  \nonumber \\
 =   i \bar{B}_+ \gamma^\mu \partial_\mu B_+ + 
i \bar{B}_- \gamma^\mu \partial_\mu B_-
- m_0 \bar{B}_+ B_+ - m_0 \bar{B}_- B_- .
\label{lag}
\end{equation}

\noindent
Alternative forms for this Lagrangian can be found in refs. \cite{TKU,TIT}.

A crucial
property of this Lagrangian  is that  the fermions
$B_+$ and $B_-$ are exactly degenerate and
have a nonzero chiral-invariant mass $m_0$. In contrast, for
usual  (Dirac) fermions chiral symmetry in the Wigner-Weyl mode 
restricts particles to be massless, hence they acquire
their mass only in the Nambu-Goldstone mode of chiral symmetry
 due to chiral symmetry breaking in the vacuum (i.e. via the coupling
with the quark condensate of the vacuum). The chiral parity doublets
have their chiral-invariant mass term already in the Wigner-Weyl mode
and this mass term has no relation with the quark condensate.

From the axial transformation law (\ref{VAD}) one can read off the
axial charge matrix, which is $\gamma_5 \sigma_1$. Hence the diagonal axial
charges of the opposite parity baryons are exactly 0, $g_+^A = g_-^A = 0$,
while the off-diagonal axial charge is 1,  $ |g_{+-}^A| = |g_{-+}^A| = 1$.
This is another crucial property that distinguishes the parity
doublets from the Dirac fermions where $g^A = 1$.
The axial vector current conservation,
$q^\mu \langle B_\pm | A_\mu |  B_\pm \rangle = 0$,
translates this axial charge matrix via the Goldberger-Treiman relation
into the $\pi B_{\pm}B_{\pm}$ coupling constants which are zero. Hence a small
(vanishing) value of the pion-baryon coupling constant taken together
with the large baryon mass would tell us that the origin of this mass is not
due to chiral symmetry breaking in the vacuum. 
An experimental verification of the smallness of the diagonal axial
charges or smallness of the pion-baryon coupling constants would be a direct
verification of the chiral symmetry restoration scenario in excited nucleons. 
It is unclear, however, how to measure experimentally these quantities.

There is  rich experimental data on strong decays of excited hadrons. It
turnes out that the  chiral restoration  implies a very strong
selection rule \cite{G4}. Namely, it predicts that if chiral symmetry is 
completely
restored in a given excited nucleon ($B$), then it cannot decay into the 
$\pi N$ channel,
 i.e. the coupling constant $f_{BN\pi}$ must vanish. This selection rule 
 is based exclusively on general properties of chiral symmetry and 
hence is model-independent.

Let us prove this selection rule.
Assume that a $\pi N$ decay of an exact chiral doublet
is possible. Then there must be a self-energy contribution
$B_\pm \rightarrow \pi N \rightarrow B_\pm$ into its mass. Then
the axial rotation  (\ref {VAD}) would require that the S-wave
$\pi N$ state transforms into the P-wave $\pi N$ state. However,
in the Nambu-Goldstone mode the axial rotations of the pion and
nucleon states are fixed - these are the nonlinear axial transformations
\cite{JPS1,JPS2}. Given these well known axial transformation
properties of the Goldstone boson and nucleon \cite{wein} it is not possible
to rotate the S-wave
$\pi N$ state  into the P-wave $\pi N$ state. Therefore, there cannot be
any $\pi N$ self-energy component in $B_\pm$. 
Hence a decay
$B_\pm \rightarrow \pi N$ is forbidden.
However, a
 decay of the exact chiral doublet into e.g. $N\rho$ or $N\pi\pi$ is not 
 forbidden.
Hence, if a state is a member of an approximate chiral multiplet, then its
decay into $N \pi$ must be strongly suppressed, 
$(f_{B N\pi}/f_{NN\pi})^2 \ll 1$. 
 
 If, in contrast, the excited baryon has no chiral partner, then its
mass, like in the nucleon case is exclusively due to chiral symmetry breaking
in the vacuum. Its axial charge should be comparable with the nucleon
axial charge. Then  nothing forbids its strong decay into $N\pi$. One then
expects that the decay coupling constant should be of the same order
as the pion-nucleon coupling constant. These two extreme cases suggest
that a magnitude of the $BN\pi$ decay constant can be used as an indicator
of the mass origin.

\begin{table}
\begin{center}
\caption{Chiral multiplets of excited nucleons.
Comment:  There
are two possibilities to assign the chiral representation:
$(1/2,0) \oplus (0,1/2)$ or $(1/2,1) \oplus (1,1/2)$ because
there is a possible chiral pair in the $\Delta$ spectrum
with the same spin with similar mass. }
\begin{tabular}{|llll|} \hline
Spin & Chiral multiplet &  Representation  & 
$(f_{B_+N\pi}/f_{NN\pi})^2 -  (f_{B_-N\pi}/f_{NN\pi})^2$ \\ \hline
1/2& $N_+(1440 ) - N_-(1535)$ & $(1/2,0) \oplus (0,1/2)$ &
   0.15 - 0.026    \\

1/2& $N_+(1710) - N_-(1650)$ & $(1/2,0) \oplus (0,1/2)$ &
 0.0030 - 0.026   \\

3/2& $N_+(1720) - N_-(1700)$ & $(1/2,0) \oplus (0,1/2)$ &
 0.023 - 0.13     \\

5/2&$N_+(1680) - N_-(1675)$ & $(1/2,0) \oplus (0,1/2)$ &
 0.18 - 0.012   \\

7/2&$N_+(?) - N_-(2190)$ &  see comment   &
  ? - 0.00053   \\

9/2&$N_+(2220) - N_-(2250)$ &
 see comment  &
 0.000022 - 0.0000020  \\

11/2&$N_+(?) - N_-(2600)$ &   see comment  &
 ? - 0.000000064   \\

\hline
\hline
3/2& $ N_-(1520)$ & no chiral partner &
   2.5     \\
\hline

\end{tabular}
\end{center}
\label{t3}
\end{table}

The decay constants
 $f_{BN\pi}$ can be extracted from the $B \rightarrow N + \pi$ decay widths,
 see e.g. \cite{CK,RB}. The pion-nucleon coupling constant is well-known,
 $f_{NN\pi} =1.0$. In Table 1 we show ratios $(f_{BN\pi}/f_{NN\pi})^2$ for
 all well-established states. It is well seen that this ratio is $\sim 0.1$
 or smaller for approximate $J=1/2,3/2,5/2$ parity doublets. For the high-spin
 states this ratio is practically vanishing. This is consistent with the
 recent demonstration of the large J-rate of chiral restoration within the only 
 known exactly solvable
 confining and chirally-symmetric model \cite{WG}.

From  Fig. 1 one can see that the only well established
excited state which has no
obvious chiral partner is $3/2^-, ~ N(1520)$. It decays
very strongly into $N \pi$, indeed. This implies that a nature of mass of
 this state is rather different compared to approximate parity doublets.
One observes a 100\% correlation
of the spectroscopic patterns with the $N \pi$ decays, as predicted
by the chiral symmetry restoration.

The Fig. 1 and the Table 1 suggest  that the lowest
approximate chiral doublet is $N(1440) - N(1535)$. If correct, the
diagonal axial charges of these states must be small. 
While it is impossible to measure these charges experimentally,
this can be done on the lattice. The axial charge of $N(1535)$ has just been
measured by Takahashi and Kunihiro and they report it to be 
surprisingly small, smaller than 0.2 \cite{TK}. Certainly lattice studies
of other states are welcome.

\section{Symmetries in excited mesons}

\begin{figure}
\begin{center}
\includegraphics[height=5cm,,clip=]{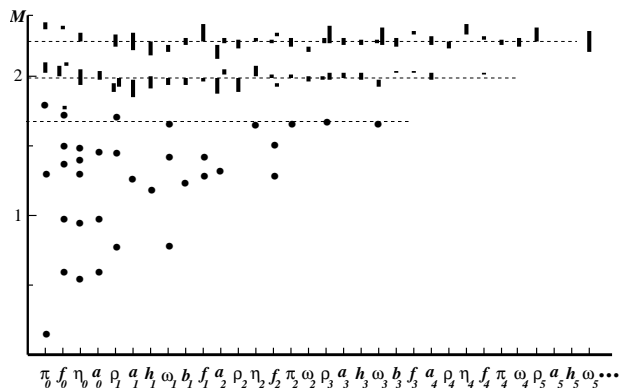}
\caption{Masses (in GeV) of the well established  states from PDG 
(circles) and 
new $\bar n n$
states   from the proton-antiproton annihilation (strips). Note
that the well-established states include $f_0(1500), f_0(1710)$, which
are the glueball and $\bar s s$ states with some mixing and hence are
irrelevant from the chiral symmetry point of view. Similar, the
 $f_0(980), a_0(980)$ mesons most probably are not $\bar n n$ states and
 also should be excluded from the consideration. The same is true for
 $\eta(1475)$, which is the $\bar s s$ state and  $\eta(1405)$ with
 the unknown nature.}
\label{lear}
\end{center}
\end{figure}

Fig. 2 shows  the spectra of the well established mesons from the PDG and
new, not yet confirmed $\bar n n$ states from the partial wave analysis 
\cite{BUGG1,BUGG2} of $\bar p p$ annihilation at LEAR (CERN). Obvious 
high symmetry of the high-lying $\bar n n$ states is seen. These data
have been analysed in ref. \cite{G5} and it turned out that the
high-lying $\bar n n$ mesons perfectly fit all possible linear
chiral multiplets of both $SU(2)_L \times SU(2)_R$ and $U(1)_A$ groups
with a few still missing states. In particular, the chiral symmetry
predicts a duplication of some of the $J > 0$ states with the given
quantum numbers, which is indeed observed in data.
If the chiral symmetry is
indeed responsible for positive-negative parity degeneracy of the states,
then there should be chiral multiplets for the high-spin states at the
levels $M \sim 2$ GeV, $M \sim 2.3$ GeV and, possibly, at $M \sim 1.7$ GeV.
These states are presently missing in refs. \cite{BUGG1,BUGG2}  and it would
be extraordinary important to find them or to reliably exclude them. Note
that such high-spin parity doublets are well seen in the nucleon spectrum -
see Fig. 1. 

The chiral and $U(1)_A$ symmetries can connect only states
with the same spin. Certainly we observe larger degeneracy, the states
with different spins are also degenerate. The large degeneracy  might be 
understood if, on top of chiral and $U(1)_A$ restorations, a principal 
quantum number $ N = n + J$ existed. 

There are suggestions in the literature to explain this large degeneracy
without resorting to chiral symmetry, assuming the $\vec J = \vec L +\vec S$
coupling scheme and that there is a principal
quantum number  $ N = n + L$, where $L$ is the {\it conserved}
orbital angular momentum
in the quark-antiquark system \cite{af,kl,sv}. This suggestion is hard to 
reconcile with the Lorentz and chiral symmetries, however \cite{GN}.

\section{Chirally symmetric and confining solvable model}

There exists only one known manifestly chirally-symmetric and confining
model in four dimensions that is solvable \cite{Orsay}, sometimes
called  Generalized Nambu and Jona-Lasinio model (GNJL). This model
can be considered as a generalization of the 1+1 dimensional
't Hooft  model, that is QCD in the large $N_c$ limit \cite{HOOFT}.
It is postulated within the GNJL model that there
exists a linear confining potential of the Coulomb type in four dimensions.
The chiral symmetry breaking 
and the properties of the Goldstone bosons have
been obtained from the solution of the Schwinger-Dyson and Bethe-Salpeter
equations  \cite{Adler:1984ri,Alkofer:1988tc,BR,BN,COT,W}. The complete
spectrum of $\bar q q$ mesons has been calculated only recently,
 in ref. \cite{WG},
which exhibits restoration of the chiral symmetry. 

\begin{figure}
\begin{center}
\includegraphics[height=5.cm]{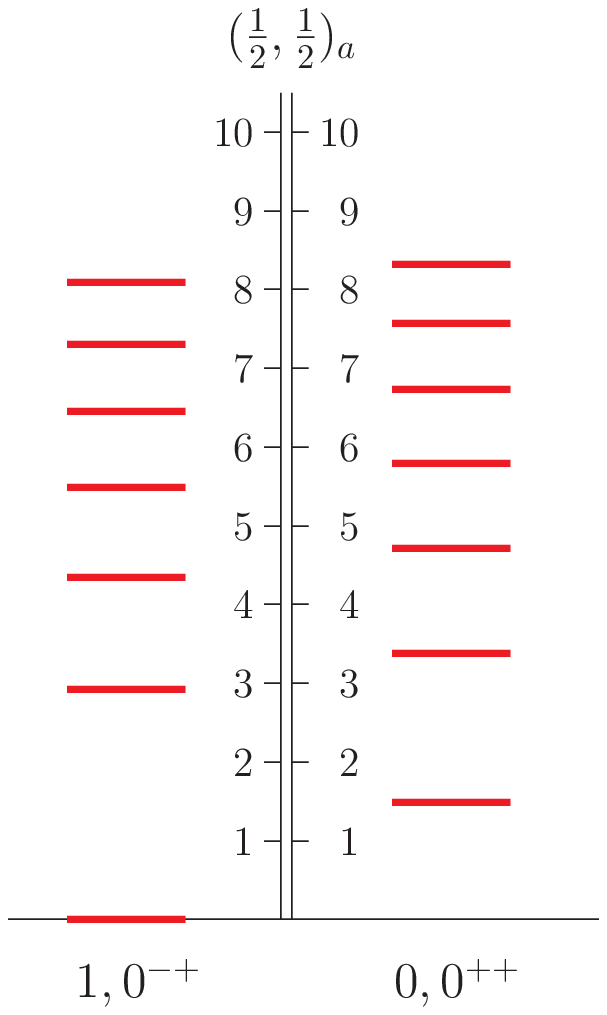}
\includegraphics[height=5.cm,clip=]{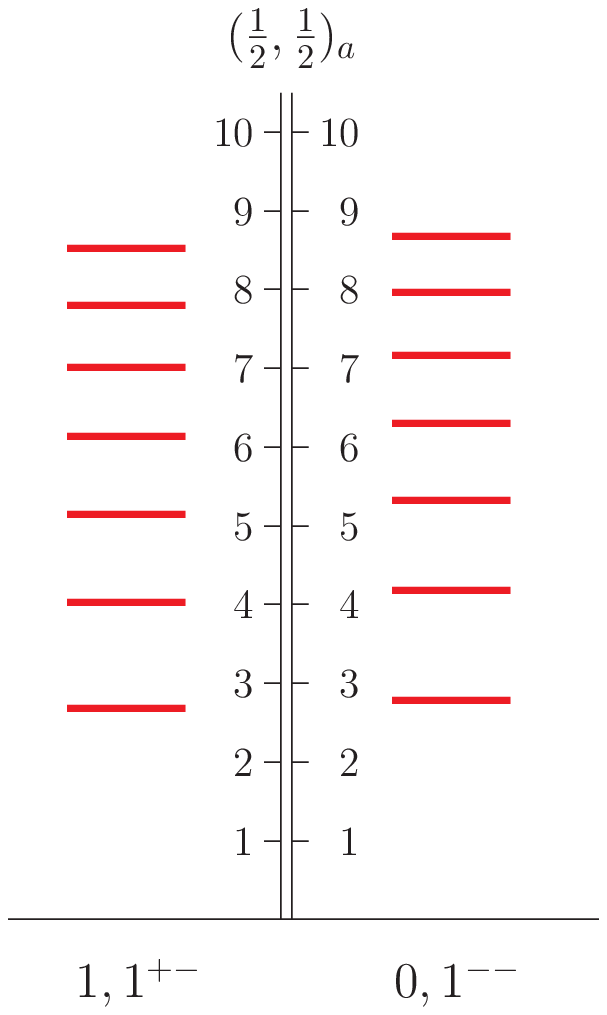}
\includegraphics[height=5.cm,clip=]{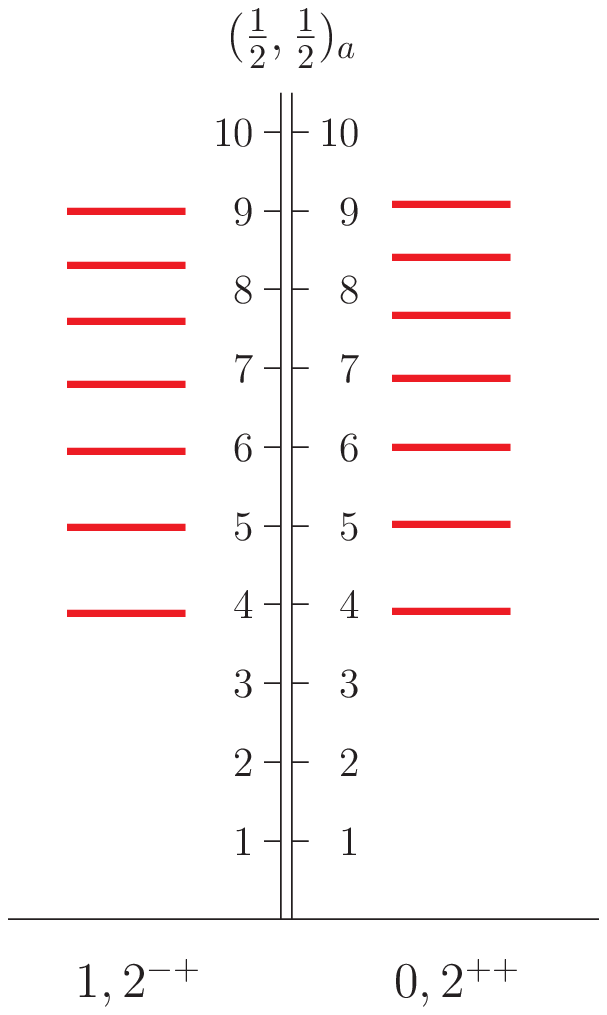}
\caption{Spectra of the $\bar q q$ mesons in the $(1/2,1/2)_a$ representations.}
\end{center}
\end{figure}

Part of the spectra is shown in Fig. 3 and a fast chiral restoration with 
increasing of $J$ is observed, while a slow rate is seen with respect
to the radial quantum number $n$. It is possible to see directly a mechanism
of the chiral restoration. The chiral symmetry breaking Lorentz-scalar
dynamical mass of quarks $M(q)$ arises via selfinteraction loops and
vanishes fast at large momenta. When one increases the spin of the hadron
$J$, or its radial quantum number $n$, one also increases the typical
momentum of valence quarks. Consequently, the chiral symmetry violating
dynamical mass of quarks becomes small and chiral symmetry gets approximately
restored. This mechanism of chiral restoration is in accord with a general
semiclassical analysis \cite{G2,G3,GNR}.

A higher degeneracy is recovered for $J \rightarrow \infty$. In this
limit all states with the same $J$ and $n$ fall into reducible representation
$[(0,1/2) \oplus (1/2,0)] \times [(0,1/2) \oplus (1/2,0)]$,
 hence the quantum loop effects
become irrelevant and all possible states with different quark
chiralities become equivalent.

\medskip
 Support of  the Austrian Science Fund through grant
P19168-N16 is  acknowledged.

\end{document}